\documentclass[twocolumn,showpacs,preprintnumbers,nofootinbib,amsmath,amssymb]{revtex4}

\usepackage{graphicx}
\usepackage{dcolumn}
\usepackage{bm}
\usepackage{epsfig}

\begin{document}

\preprint{DESY~14-219\hspace{14.2cm} ISSN~0418-9833}
\preprint{November 2014\hspace{16.3cm}}

\title{\boldmath $\eta_c$ production at the LHC challenges nonrelativistic-QCD
factorization}

\author{Mathias Butenschoen}
\author{Zhi-Guo He}
\author{Bernd A. Kniehl}
\affiliation{{II.} Institut f\"ur Theoretische Physik, Universit\"at Hamburg,
Luruper Chaussee 149, 22761 Hamburg, Germany}

\date{\today}

\begin{abstract}
We analyze the first measurement of $\eta_c$ production, performed by the LHCb
Collaboration, in the nonrelativistic-QCD (NRQCD) factorization framework at
next-to-leading order (NLO) in the strong-coupling constant $\alpha_s$ and the
relative velocity $v$ of the bound quarks including the feeddown from $h_c$
mesons.
Converting the long-distance matrix elements (LDMEs) extracted by various
groups from $J/\psi$ yield and polarization data to the $\eta_c$ case using
heavy-quark spin symmetry, we find that the resulting NLO NRQCD predictions
greatly overshoot the LHCb data, while the color-singlet model provides an
excellent description.
\end{abstract}

\maketitle

Despite concerted experimental and theoretical efforts ever since the
discovery of the $J/\psi$ meson in the November revolution of 1974 (The Nobel
Prize in Physics 1976), the genuine mechanism underlying the production and
decay of heavy quarkonia, which are QCD bound states of a heavy quark $Q=c,b$
and its antiparticle $\overline{Q}$, has remained mysterious.
The effective quantum field theory of nonrelativistic QCD (NRQCD)
\cite{Caswell:1985ui} endowed with an appropriate factorization theorem
\cite{Bodwin:1994jh}
arguably constitutes the most probable
candidate theory at the present time.
This implies a separation of process-dependent short-distance coefficients
(SDCs), to be calculated perturbatively as expansions in the strong-coupling
constant $\alpha_s$, from supposedly universal long-distance matrix elements
(LDMEs), to be extracted from experiment.
The relative importance of the latter is subject to velocity scaling rules
\cite{Lepage:1992tx}, which predict each of the LDMEs to scale with a definite
power of the heavy-quark velocity $v$.
In this way, the theoretical predictions are organized as double expansions in
$\alpha_s$ and $v$.
A crucial feature of this formalism is that the $Q\overline{Q}$ pair can at
short distances be produced in any Fock state $n={}^{2S+1}L_J^{[a]}$ with
definite spin $S$, orbital angular momentum $L$, total angular momentum $J$,
and color multiplicity $a=1,8$.
In this way, it complements the color-singlet (CS) model (CSM), which only
includes the very ${}^{2S+1}L_J^{[1]}$ state of the physical quarkonium, and
thus cures a severe conceptual shortcoming of the latter, namely the existence
of uncanceled infrared (IR) singularities beyond $L=0$.
However, the CSM does provide IR-finite NLO predictions for $S$-wave charmonia,
such as the $\eta_c$ and $J/\psi$ mesons considered here.

Despite its theoretical rigor, NRQCD factorization has reached the crossroads
in the $J/\psi$ case.
While a global fit \cite{Butenschoen:2011yh} to the $J/\psi$ yields measured in
hadroproduction, photoproduction, $\gamma\gamma$ scattering, and $e^+e^-$
annihilation successfully pins down the leading color-octet (CO) LDMEs,
$\langle{\cal O}^{J/\psi}(^1\!S_0^{[8]})\rangle$,
$\langle{\cal O}^{J/\psi}(^3\!S_1^{[8]})\rangle$, and
$\langle{\cal O}^{J/\psi}(^3\!P_0^{[8]})\rangle$,
in compliance with the velocity scaling rules, the resulting predictions for
$J/\psi$ polarization in hadroproduction are in striking disagreement
with measurements at the Fermilab Tevatron and the CERN LHC
\cite{Butenschoen:2012px}.
Vice versa, fits to data on $J/\psi$ yield and polarization in hadroproduction
work reasonably well \cite{Chao:2012iv,Gong:2012ug,Bodwin:2014gia}, but
hopelessly fail in comparisons to the world's data from other than hadronic
collisions \cite{Butenschoen:2012qr}, with transverse momenta up to
$p_T=10$~GeV.

Very recently, the LHCb Collaboration measured, for the first time, the prompt
$\eta_c$ yield, via $\eta_c\to p\bar{p}$ decays \cite{Aaij:2014bga}.
The data were taken at center-of-mass energies $\sqrt{s}=7$ and 8~TeV in the
forward rapidity range $2.0<y<4.5$ in bins of $p_T$.
This provides a tantalizing new opportunity to further test NRQCD factorization
and, hopefully, to also shed light on the $J/\psi$ polarization puzzle, the
more so as the $\eta_c$ meson is the spin-singlet partner of the $J/\psi$
meson, which implies that the LDMEs of the two are related by heavy-quark spin
symmetry (HQSS), one of the pillars of NRQCD factorization.
The dominant feeddown contribution is due to the radiative decay
$h_c\to\eta_c\gamma$.
The leading CS and CO Fock states of direct $\eta_c$ ($h_c$) production are
${}^1\!S_0^{[1]}$ at $\mathcal{O}(v^3)$ and
${}^1\!S_0^{[8]}$, ${}^3\!S_1^{[8]}$, and ${}^1\!P_1^{[8]}$
at $\mathcal{O}(v^7)$
(${}^1\!P_1^{[1]}$ and ${}^1\!S_0^{[8]}$ at $\mathcal{O}(v^5)$).

So far, only incomplete LO calculations were carried out for direct $\eta_c$
production, excluding the ${}^1\!S_0^{[8]}$ contribution \cite{Mathews:1998nk}.
For the reasons explained above, it is an urgent matter of general interest to
provide a full-fledged NRQCD analysis of prompt $\eta_c$ hadroproduction, at
NLO both in $\alpha_s$ and $v$, and this is the very purpose of this Letter.
From the $J/\psi$ case, where such systematic investigations already exist
\cite{Butenschoen:2011yh,Chao:2012iv,Gong:2012ug,Bodwin:2014gia}, we know (i)
that $\mathcal{O}(\alpha_s)$ corrections may be sizable, especially in the
${}^3\!P_J^{[8]}$ channels, (ii) that $\mathcal{O}(v^2)$ corrections may be
non-negligible \cite{Xu:2012am,He:2014rc}, and (iii) that feeddown
contributions to prompt production may be substantial, reaching 20--30\% in the
$\chi_{cJ}$ case \cite{Abe:1997yz,LHCb:2012af,Gong:2012ug}.

We work in the collinear parton model of QCD implemented in the
fixed-flavor-number scheme with $n_f=3$ quark flavors active in the colliding
protons, which are represented by parton density functions (PDFs) evaluated at
factorization scale $\mu_f$.
At NLO in NRQCD, the relevant partonic cross sections are given by
\begin{eqnarray}
&&d\sigma_\mathrm{prompt}^{\eta_c}=
\sum_{n={}^1\!S_0^{[1]},{}^1\!S_0^{[8]},{}^3\!S_1^{[8]}, {}^1\!P_1^{[8]}} \Big[
d\sigma^{c\overline{c}[n]} \langle {\cal O}^{\eta_c}(n)\rangle
\nonumber\\
&&{}+d\sigma_{v^2}^{c\overline{c}[n]} \langle {\cal P}^{\eta_c}(n) \rangle \Big]
+\sum_{n={}^1\!P_1^{[1]},{}^1\!S_0^{[8]}} \Big[
d\sigma^{c\overline{c}[n]}\langle {\cal O}^{h_c}(n)\rangle
\nonumber\\
 &&{}+ d\sigma_{v^2}^{c\overline{c}[n]} \langle {\cal P}^{h_c}(n) \rangle \Big]
{\cal B}(h_c\to\eta_c\gamma),
\label{eq:etacfact}
\end{eqnarray}
where $d\sigma^{c\overline{c}[n]}$ are the Born SDCs including their
$\mathcal{O}(\alpha_s)$ corrections, $d\sigma_{v^2}^{c\overline{c}[n]}$ contain
their $\mathcal{O}(v^2)$ corrections, and
$\langle {\cal Q}^h(n)\rangle$ with $\mathcal{Q}=\mathcal{O},\mathcal{P}$ and
$h=\eta_c,h_c$ are the appropriate LDMEs.
We approximately account for the mass difference between the $\eta_c$ and $h_c$
mesons by substituting $p_T\to p_T m_{h_c}/m_{\eta_c}$ in the $h_c$ SDCs.
The definitions of the $\mathcal{O}$ and $\mathcal{P}$ operators for $S$-wave
states and the $\mathcal{O}$ operator for the $P$-wave states may be found in
Refs.~\cite{Bodwin:1994jh,He:2014rc}.
Analogously, we define the $\mathcal{P}$ operators of the relevant $P$-wave
states as
\begin{eqnarray}
\mathcal{P}^{\eta_c}(^{1}P_{1}^{[8]})=\chi^{\dagger}\left(-\frac{i}{2}
\overleftrightarrow{\boldsymbol{D}^{j}}\right)T^{a}\psi
(a^{\dagger}_{\eta_c}a_{\eta_c})
\nonumber \\
\times \psi^{\dagger}T^{a}\left(-\frac{i}{2}
\overleftrightarrow{\boldsymbol{D}^{j}}\right)\left(-\frac{i}{2}
\overleftrightarrow{\boldsymbol{D}}\right)^{2}\chi+{\rm H.c.},
\nonumber\\
\mathcal{P}^{h_c}(^{1}P_{1}^{[1]})=\chi^{\dagger}\left(-\frac{i}{2}
\overleftrightarrow{\boldsymbol{D}^{j}}\right)\psi(a^{\dagger}_{h_c}a_{h_c})
\nonumber \\
\times\psi^{\dagger}\left(-\frac{i}{2}\overleftrightarrow{\boldsymbol{D}^{j}}\right)
\left(-\frac{i}{2}\overleftrightarrow{\boldsymbol{D}}\right)^{2}\chi+{\rm H.c.}.
\end{eqnarray}
The HQSS relationships between the $\eta_c$ and $J/\psi$ ($h_c$ and
$\chi_{c0}$) LDMEs, which are exact through $\mathcal{O}(v^2)$, read
\cite{Bodwin:1994jh}:
\begin{eqnarray}
 \langle {\cal Q}^{\eta_c}({}^1\!S_0^{[1]}/\,{}^1\!S_0^{[8]}) \rangle&=& \frac{1}{3}
 \langle {\cal Q}^{J/\psi}({}^3\!S_1^{[1]}/\,{}^3\!S_1^{[8]}) \rangle
\nonumber\\
 \langle {\cal Q}^{\eta_c}({}^3\!S_1^{[8]}) \rangle&=&
 \langle {\cal Q}^{J/\psi}({}^1\!S_0^{[8]}) \rangle
\nonumber\\
 \langle {\cal Q}^{\eta_c}({}^1\!P_1^{[8]}) \rangle&=& 3
 \langle {\cal Q}^{J/\psi}({}^3\!P_0^{[8]}) \rangle
\nonumber\\
 \langle {\cal Q}^{h_c}({}^1\!P_1^{[1]}/\,{}^1\!S_0^{[8]}) \rangle&=& 3
 \langle {\cal Q}^{\chi_{c0}}({}^3\!P_0^{[1]}/\,{}^3\!S_1^{[8]}) \rangle.
\label{eq:HQSSR}
\end{eqnarray}

At ${\cal O}(\alpha_s)$, $\langle {\cal O}({}^1\!S_0^{[1/8]})\rangle$ turn out to
be proportional to
$(1/\epsilon_\mathrm{UV}-1/\epsilon_\mathrm{IR})
\langle {\cal O}({}^1\!P_1^{[1/8]}) \rangle$,
where the poles in $\epsilon=2-d/2$, with $d$ being the space-time dimension in
dimensional regularization, are of ultraviolet (UV) or IR origin.
After appropriate $\overline{\mathrm{MS}}$ operator renormalization, the
renormalized free-quark LDMEs pertaining to the NRQCD matching procedure for
calculating the SDCs are given to $\mathcal{O}(\alpha_s)$ by
\begin{eqnarray}
&&\hspace{-5mm}\langle {\cal O}^h({}^1\!S_0^{[8]}) \rangle(\mu_\lambda) =  \langle {\cal O}^h({}^1\!S_0^{[8]}) \rangle_0
- \frac{4\alpha_s(\mu_\lambda)}{3\pi m^2} \left(\frac{4 \pi \mu^2}{\mu_\lambda^2}e^{-\gamma_E}\right)^\epsilon
\nonumber \\
&&\hspace{-5mm}{}\times \frac{1}{\epsilon_\mathrm{IR}} \left[ \frac{C_F}{2C_A} \langle {\cal O}^h({}^1\!P_1^{[1]}) \rangle
+ \left(\frac{C_A}{4}-\frac{1}{C_A}\right) \langle {\cal O}^h({}^1\!P_1^{[8]}) \rangle \right]
\nonumber\\
&&\hspace{-5mm}\langle {\cal O}^h({}^1\!S_0^{[1]}) \rangle(\mu_\lambda) =  \langle {\cal O}^h({}^1\!S_0^{[1]}) \rangle_0 -
\frac{4\alpha_s(\mu_\lambda)}{3\pi m^2} \left(\frac{4 \pi \mu^2}{\mu_\lambda^2}e^{-\gamma_E}\right)^\epsilon 
\nonumber\\
&&\hspace{-5mm}{}\times \frac{1}{\epsilon_\mathrm{IR}} \frac{1}{2C_A} \langle {\cal O}^h({}^1\!P_1^{[8]}) \rangle,
\label{eq:NLOLDMEs}
\end{eqnarray}
where $\mu_\lambda$ and $\mu_r$ are the NRQCD and QCD renormalization scales,
respectively, and $\langle {\cal O}^h(n)\rangle_0$ are the tree-level LDMEs.
The IR poles in Eq.~(\ref{eq:NLOLDMEs}) match otherwise uncanceled IR poles
produced by the real radiative corrections to the $P$-wave SDCs.
The $\mu_\lambda$ dependences of the renormalized LDMEs are then determined by
solving
$\mu_\lambda \frac{d}{d\mu_\lambda} \langle {\cal O}^h(n) \rangle_0=0$
\cite{Klasen:2004tz}.
We do not need to consider NLO corrections to $P$-wave LDMEs, since they are
proportional to operators beyond $\mathcal{O}(v^2)$.

We calculate the $\mathcal{O}(\alpha_s)$ and $\mathcal{O}(v^2)$ corrections to
the SDCs using the techniques developed in
Refs.~\cite{Butenschoen:2009zy,Butenschoen:2010rq,He:2014rc}.
The $\mathcal{O}(\alpha_s)$ corrections to ${}^1\!P_1^{[1]}$ state
hadroproduction have only recently been calculated in Ref.~\cite{Wang:2014vsa}.
We can reproduce the results therein within the uncertainties expected from the
phase-space-slicing method.
We can trace the only significant difference to the variation of $\mu_\lambda$
about its default value, which was executed in Ref.~\cite{Wang:2014vsa} only
in the SDCs, where it is induced via Eq.~(\ref{eq:NLOLDMEs}), but not in the
LDMEs.
The $\mathcal{O}(\alpha_s)$ corrections to the ${}^1\!S_0^{[1]}$ and
${}^1\!P_1^{[8]}$ SDCs as well as the $\mathcal{O}(v^2)$ corrections to the
${}^1\!S_0^{[1]}$, ${}^1\!P_1^{[1]}$, and ${}^1\!P_1^{[8]}$ SDCs are calculated
here for the first time.

\begin{table*}
\begin{tabular}{c|c|c|c|c}
& Butensch\"on, & Chao, Ma, Shao, & Gong, Wan, & Bodwin, Chung,\\
& Kniehl \cite{Butenschoen:2011yh} & Wang, Zhang \cite{Chao:2012iv}
&  Wang, Zhang \cite{Gong:2012ug} & Kim, Lee \cite{Bodwin:2014gia}\\
\hline
$\langle {\cal O}^{J/\psi}({}^3\!S_1^{[1]}) \rangle /\mbox{GeV}^3 $ &
$1.32$ &
$1.16$ &
$1.16$
\\
$\langle {\cal O}^{J/\psi}({}^1\!S_0^{[8]}) \rangle /\mbox{GeV}^3$ &
$0.0304\pm0.0035$ &
$0.089\pm0.0098$ &
 $0.097\pm0.009$&
$0.099\pm0.022$
\\
$\langle {\cal O}^{J/\psi}({}^3\!S_1^{[8]}) \rangle /\mbox{GeV}^3$ &
$0.0016\pm0.0005$ &
$0.0030\pm0.012$&
$-0.0046\pm0.0013$&
$0.011\pm0.010$
\\
$\langle {\cal O}^{J/\psi}({}^3\!P_0^{[8]}) \rangle /\mbox{GeV}^5$ &
$-0.0091\pm0.0016$ &
 $0.0126\pm0.0047$&
$-0.0214\pm0.0056$ &
$0.011\pm0.010$
\\
\hline
$\langle {\cal O}^{\chi_0}({}^3\!P_0^{[1]}) \rangle /\mbox{GeV}^5$ & &
& $0.107$
\\
$\langle {\cal O}^{\chi_0}({}^3\!S_1^{[8]}) \rangle /\mbox{GeV}^3$ & &
& $0.0022\pm0.0005$
\end{tabular}
\caption{\label{tab:usedLDMEs}
Sets of $J/\psi$ and $\chi_{c0}$ LDMEs determined in
Refs.~\cite{Butenschoen:2011yh,Chao:2012iv,Gong:2012ug,Bodwin:2014gia}.}
\end{table*}

In our numerical analysis, we adopt the values $m_{\eta_c}=2983.6$~GeV,
$m_{h_c}=3525.38$~GeV, and $\mathrm{Br}(h_c\to\eta_c\gamma)=51\%$ from
Ref.~\cite{Agashe:2014kda}, take the charm-quark mass, which we renormalize
according to the on-shell scheme, to be $m_c=1.5$~GeV, and use the one-loop
(two-loop) formula for $\alpha_s^{(n_f)}(\mu_r)$ with $n_f=4$ at LO (NLO).
As for the proton PDFs, we use the CTEQ6L1 (CTEQ6M) set \cite{Pumplin:2002vw}
at LO (NLO), which comes with an asymptotic scale parameter of
$\Lambda_\mathrm{QCD}^{(4)}=215$~MeV (326~MeV).
Our default scale choices are $\mu_\lambda=m_c$ and $\mu_r=\mu_f=m_T$ with
$m_T=\sqrt{p_T^2+4m_c^2}$ being the charmonium's transverse mass.
We in turn adopt two approaches to determine the $\eta_c$ and $h_c$ LDMEs.
In the first one, we obtain them via Eq.~(\ref{eq:HQSSR}) from the $J/\psi$ and
$\chi_{c0}$ LDME sets determined at NLO, but ignoring relativistic corrections,
by four different groups
\cite{Butenschoen:2011yh,Chao:2012iv,Gong:2012ug,Bodwin:2014gia} from different
selections of $J/\psi$ and $\chi_{c0}$ production data (see
Table~\ref{tab:usedLDMEs}).
In those cases where no $\chi_{cJ}$ or CS $J/\psi$ LDMEs are available, we
omit the corresponding contributions.
The observation that direct $\eta_c$ production almost exclusively proceeds via
the ${}^3\!S_1^{[8]}$ channel will provide a retroactive justification for that.
\begin{figure*}
\centering
\includegraphics[width=8.8cm]{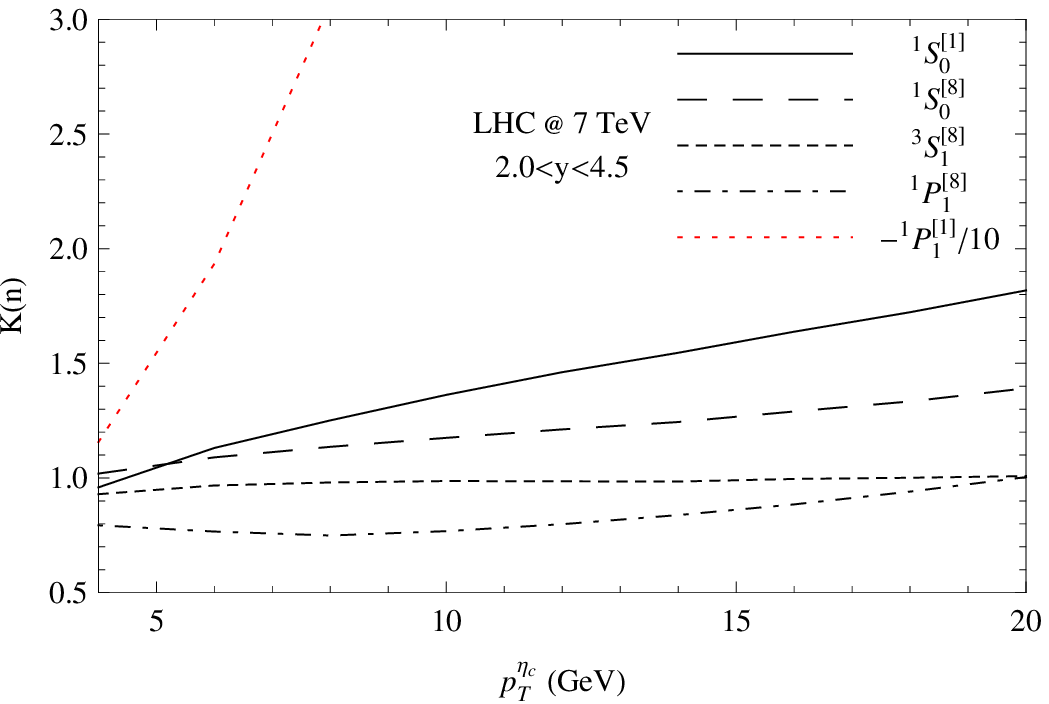}
\includegraphics[width=8.8cm]{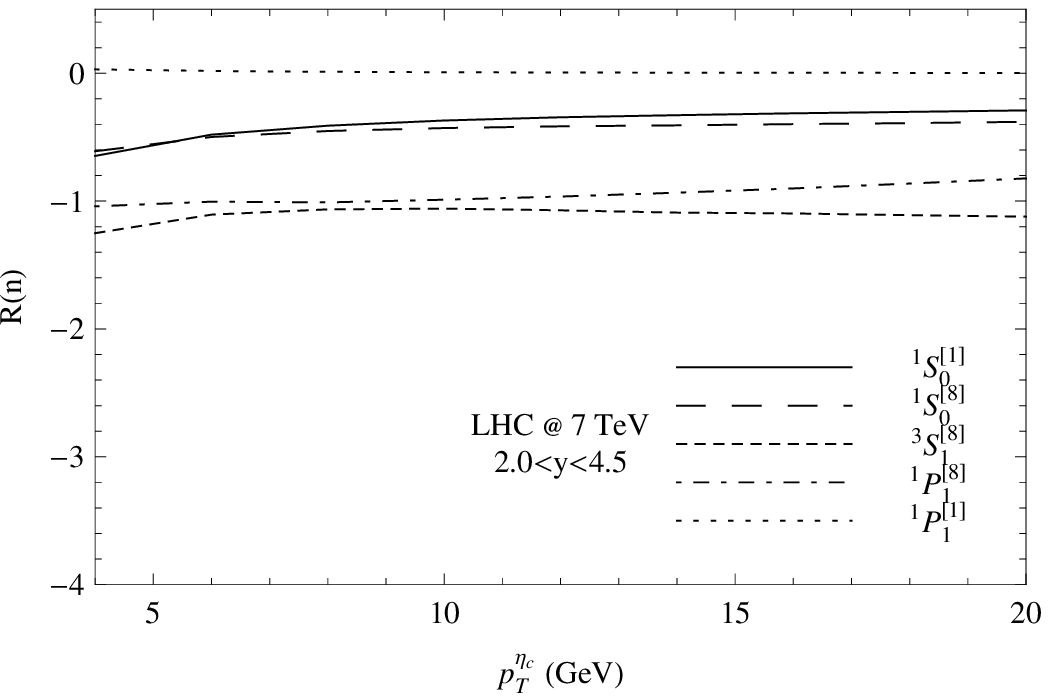}
\caption{\label{fig:kfactors}
Ratios $K(n)=d\sigma_\mathrm{NLO}^{c\bar{c}[n]}/d\sigma_\mathrm{LO}^{c\bar{c}[n]}$
measuring the $\mathcal{O}(\alpha_s)$ corrections to the SDCs as functions of
$p_T^{\eta_c}$ (left panel).
Ratios $R(n)=d\sigma_{v^2}^{c\bar{c}[n]}m_c^2/d\sigma_\mathrm{NLO}^{c\bar{c}[n]}$
measuring the $\mathcal{O}(v^2)$ corrections to the SDCs as functions of
$p_T^{\eta_c}$ (right panel).
The results for $n={}^1\!P_1^{[1]}$ refer to $h_c$ production and are evaluated at
$p_T^{h_c}=p_t^{\eta_c}m_{h_c}/m_{\eta_c}$.
The results for $n={}^1\!S_{0}^{[8]}$ in $h_c$ production are not shown, but
may be obtained from those in $\eta_c$ production by rescaling as for
$n={}^1\!P_1^{[1]}$.
Red color (minus sign in the legend) indicates negative values.}
\end{figure*}

In Fig.~\ref{fig:kfactors}, we analyze the $\mathcal{O}(\alpha_s)$ and
$\mathcal{O}(v^2)$ corrections to the contributing SDCs for unit LDMEs.
We note that the $\mathcal{O}(\alpha_s)$ corrections turn the $^1\!P_1^{[1]}$ 
SDC negative, a feature familiar, for example, from the $^3\!P_J^{[8]}$ SDC of
direct $J/\psi$ hadroproduction \cite{Butenschoen:2010rq}.
However, the $^1\!P_1^{[8]}$ SDC stays positive also after including the
${\cal O}(\alpha_s)$ corrections.
As for the $\mathcal{O}(v^2)$ corrections, we observe that the ratios 
$R(n)=d\sigma_{v^2}^{c\bar{c}[n]}m_c^2/d\sigma_\mathrm{NLO}^{c\bar{c}[n]}$ are almost
independent of $p_T$ and of order unity for all $n$, except for $^1P_{1}^{[1]}$,
which confirms that the relativistic corrections are actually of relative order
$\mathcal{O}(v^2)$.

\begin{figure*}
\centering
\includegraphics[width=4.4cm]{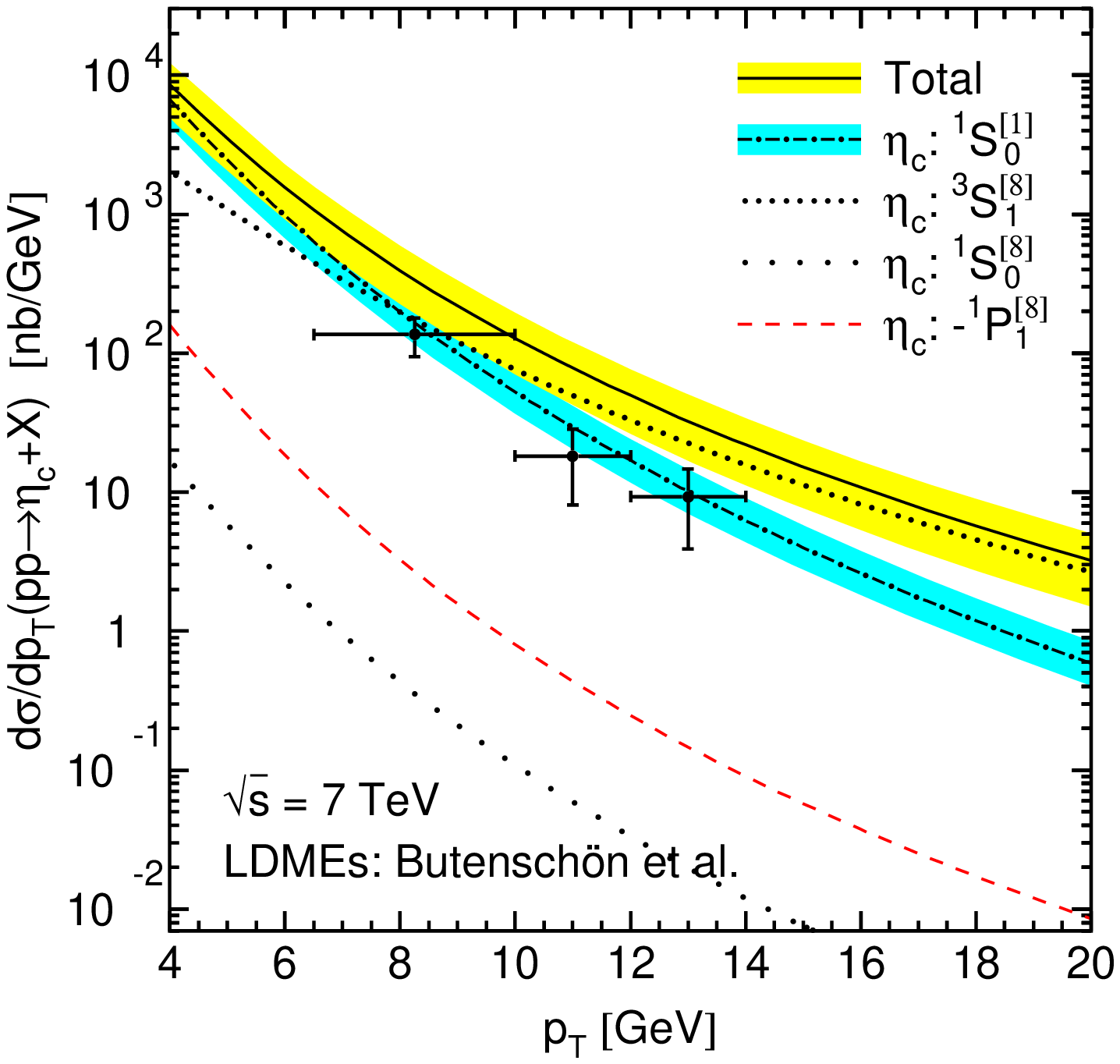}
\includegraphics[width=4.4cm]{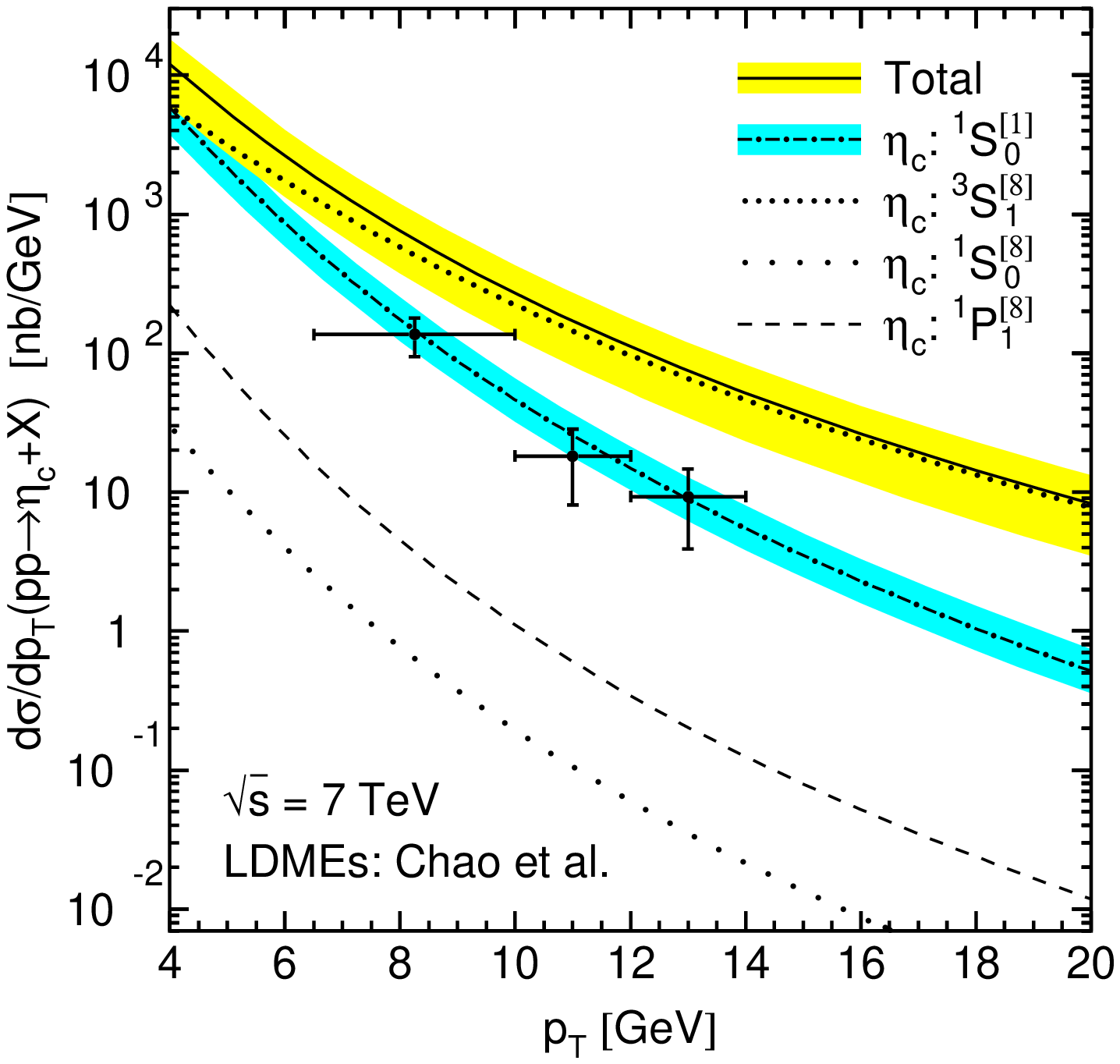}
\includegraphics[width=4.4cm]{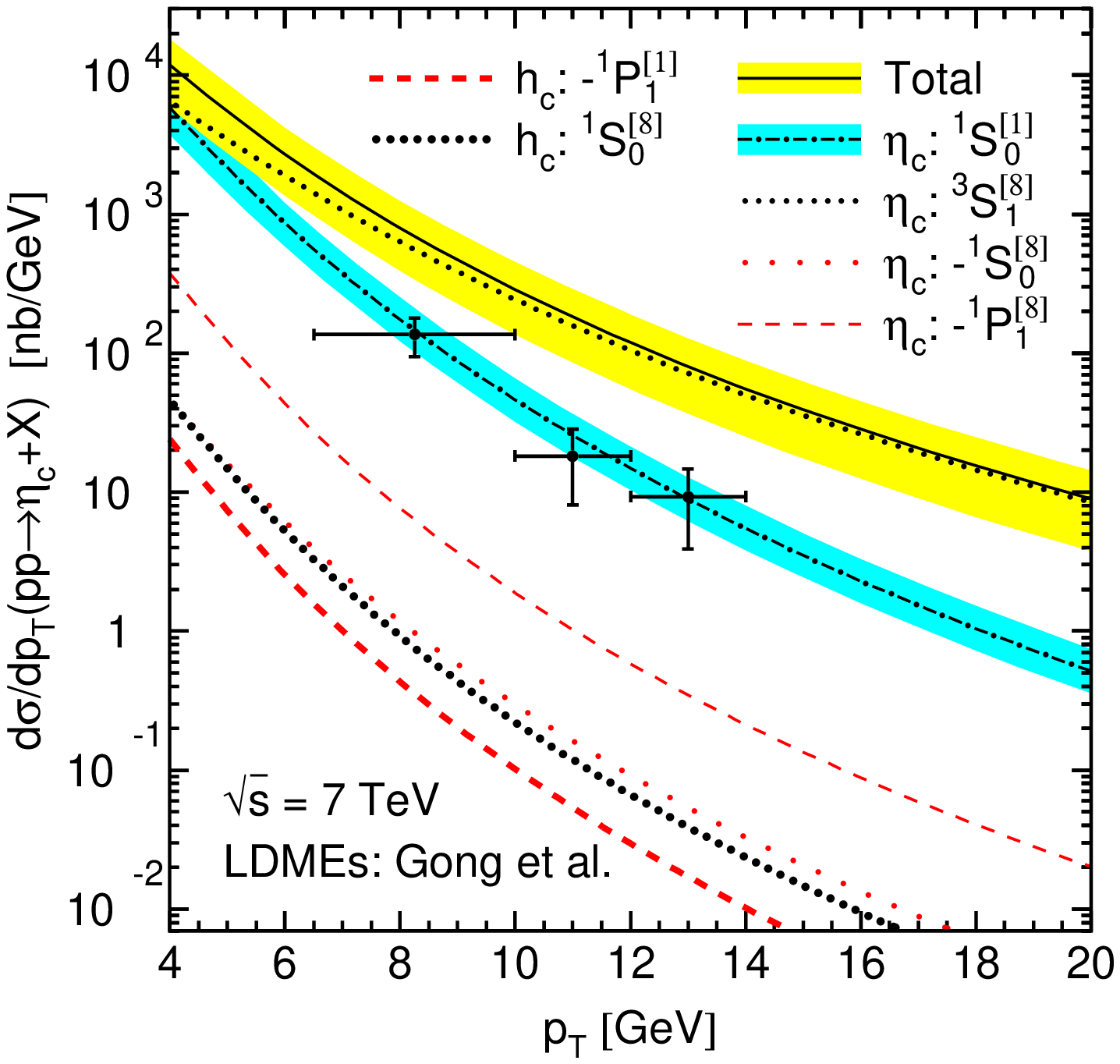}
\includegraphics[width=4.4cm]{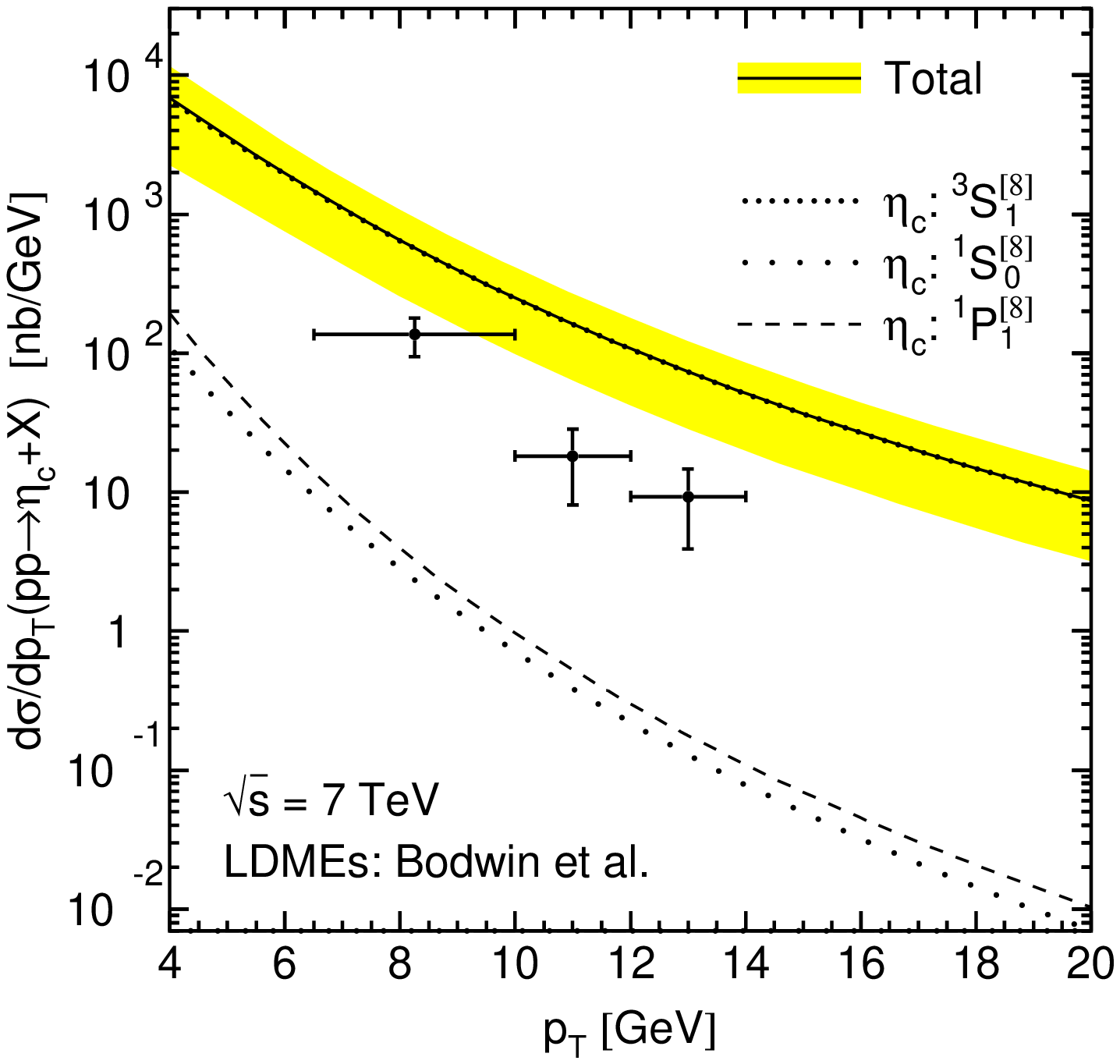}

\vspace{2pt}
\includegraphics[width=4.4cm]{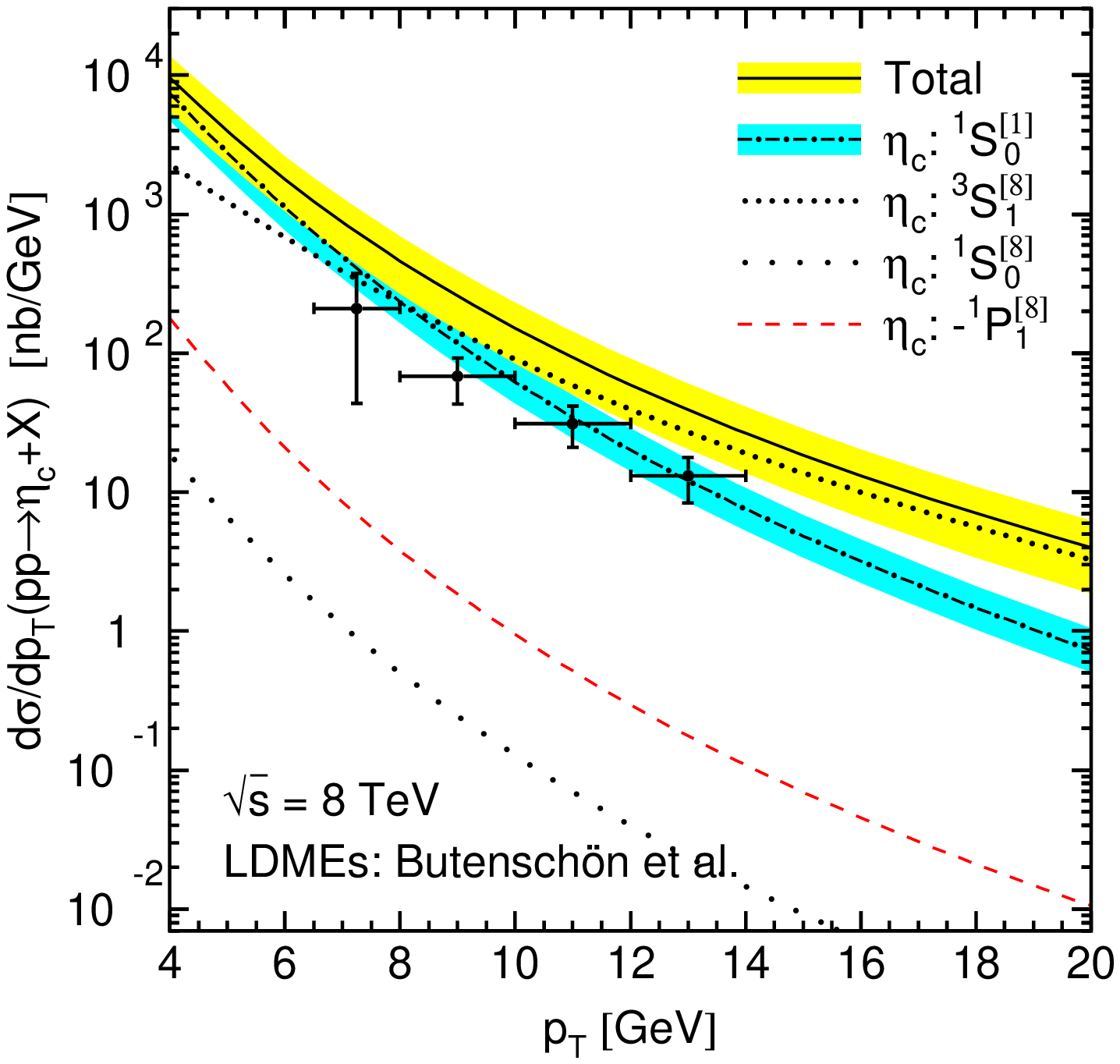}
\includegraphics[width=4.4cm]{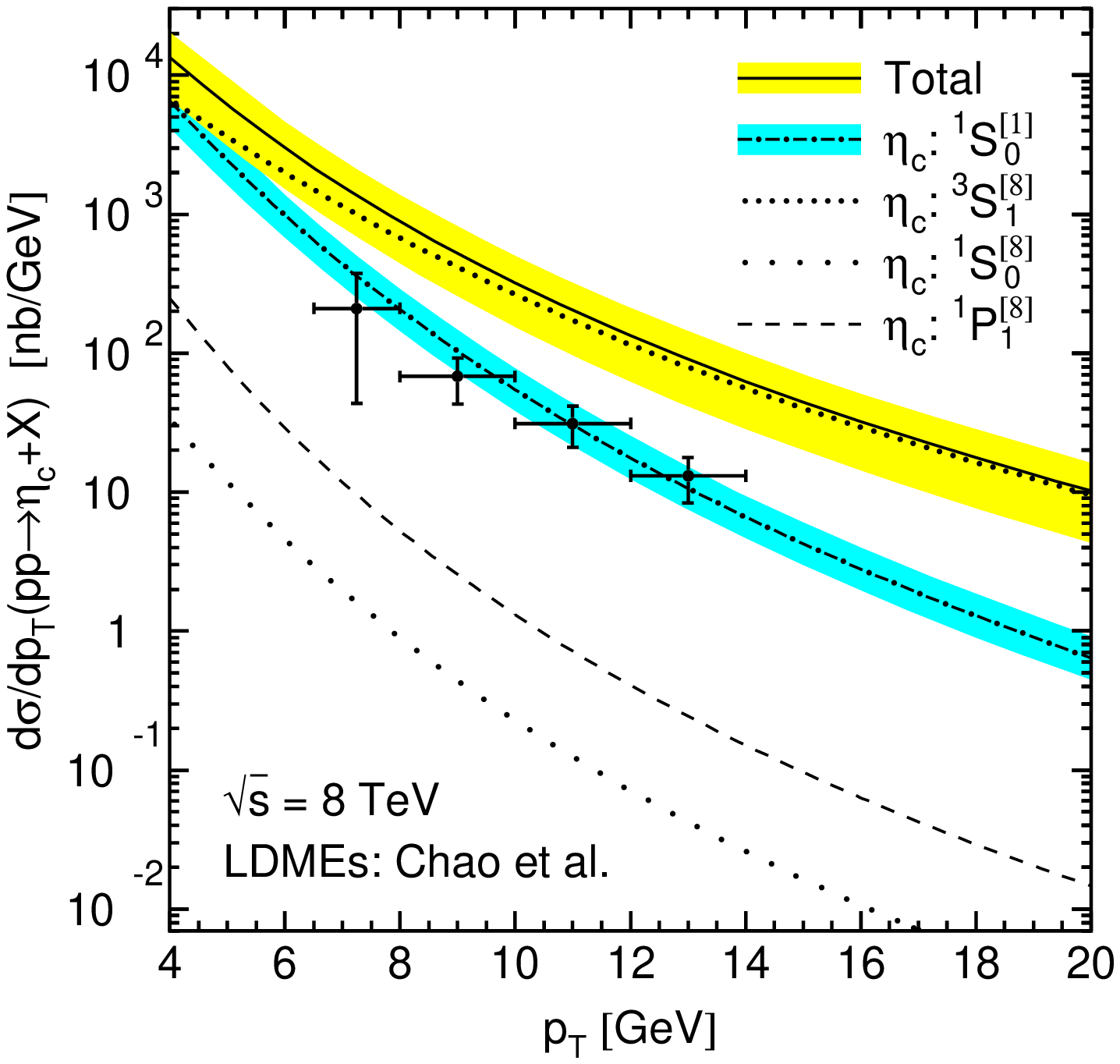}
\includegraphics[width=4.4cm]{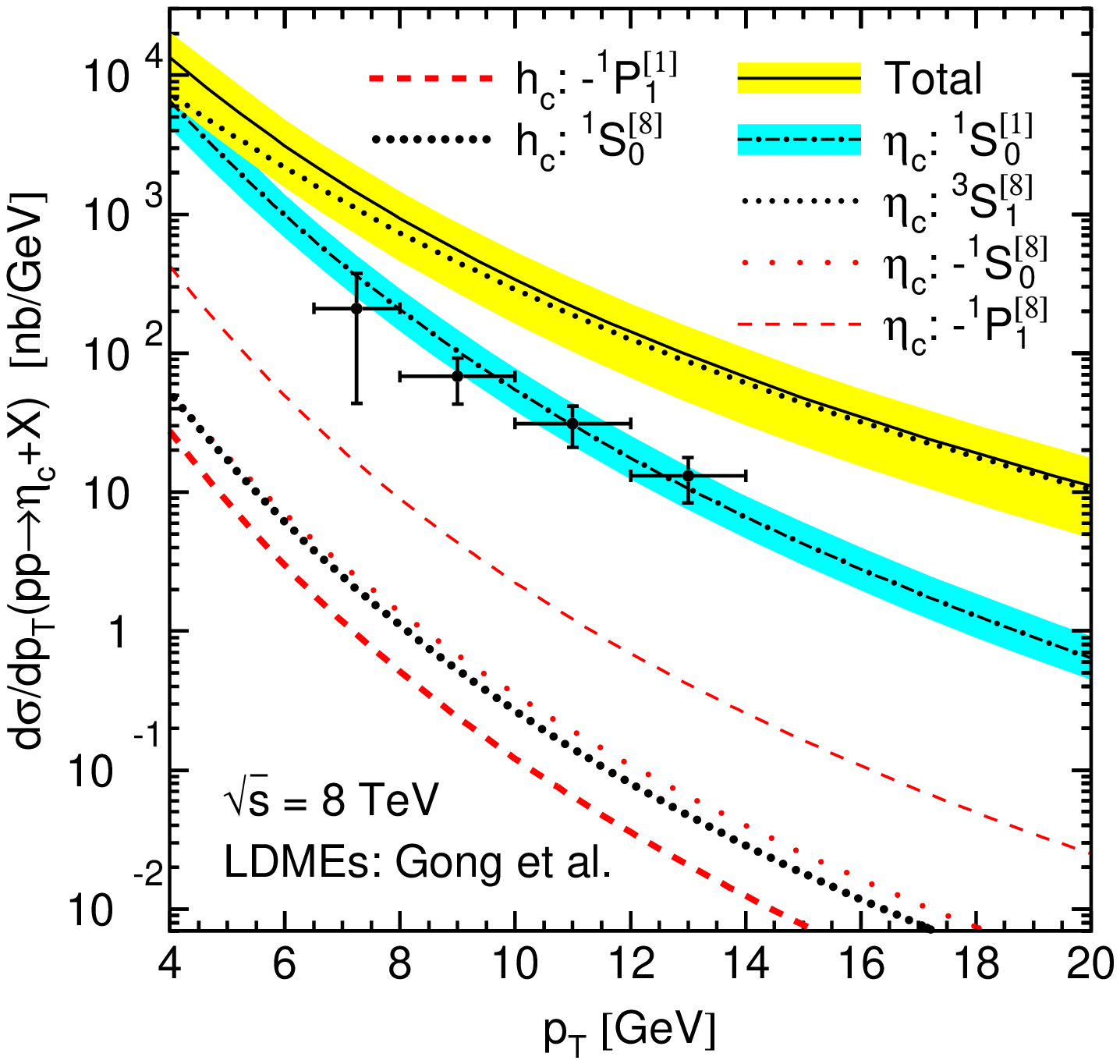}
\includegraphics[width=4.4cm]{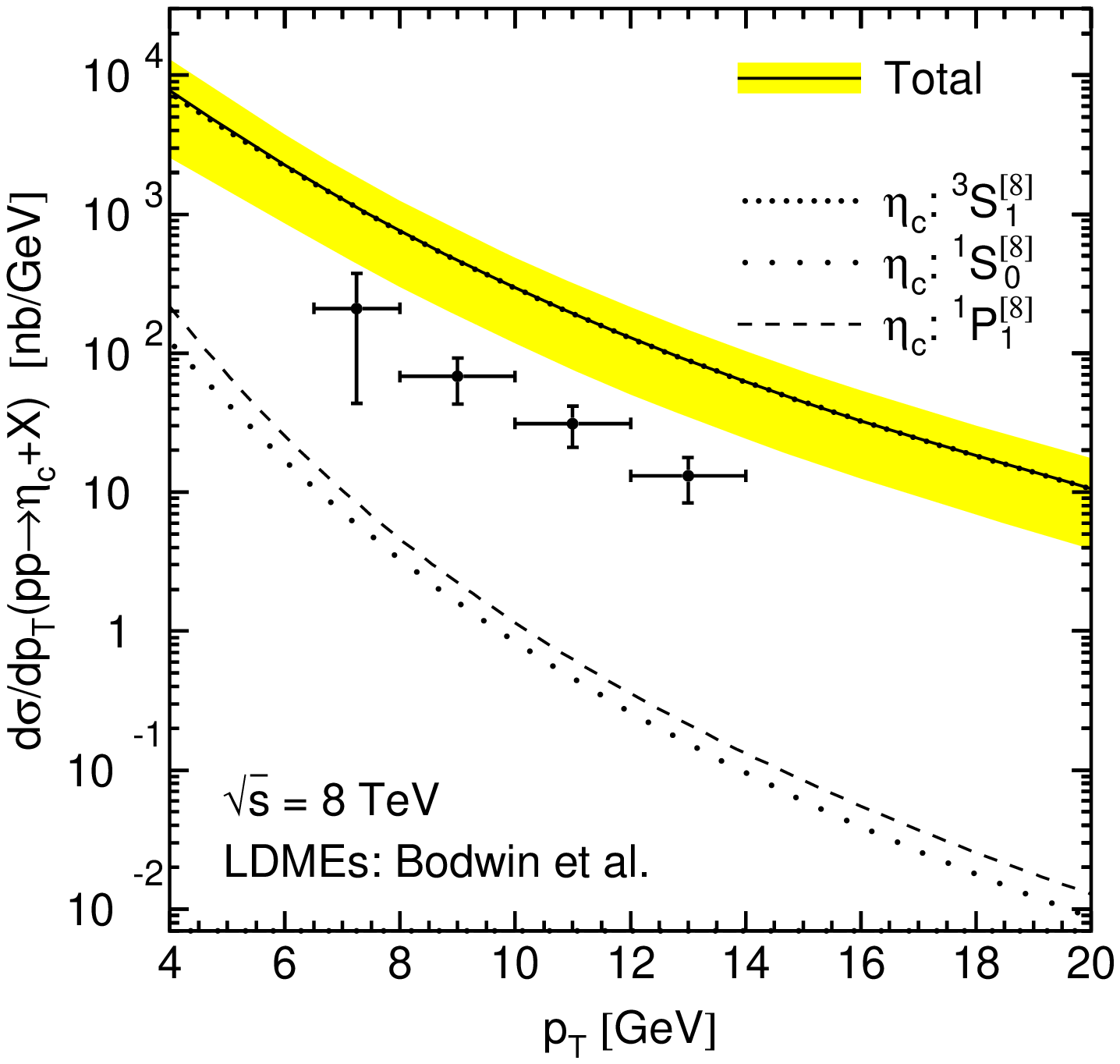}
\caption{\label{fig:etacplots}
The LHCb \cite{Aaij:2014bga} measurements of $d\sigma/dp_T$ for prompt $\eta_c$
hadroproduction at $\sqrt{s}=7$~TeV (upper panel) and 8~TeV (lower panel) are
compared with the default predictions of NRQCD (solid lines) and the CSM
(dot-dashed lines) at NLO, but without relativistic corrections, evaluated with
the four LDME sets in Table~\ref{tab:usedLDMEs}.
The theoretical errors as explained in the text are indicated by the yellow and
blue bands, respectively.
For comparison, also the default contributions due to the individual Fock
states are shown.
Red color (minus sign in the legend) indicates negative values.}
\end{figure*}

In Fig.~\ref{fig:etacplots}, the LHCb data \cite{Aaij:2014bga} are compared
with the NRQCD and CSM default predictions including $\mathcal{O}(\alpha_s)$
but excluding $\mathcal{O}(v^2)$ corrections, evaluated with the four LDME sets
in Table~\ref{tab:usedLDMEs}.
The error bands shown there are obtained by adding three theoretical errors in
quadrature.
The first is due to unknown corrections beyond $\mathcal{O}(\alpha_s)$, which
are estimated by varying $\mu_\lambda$, $\mu_r$, and $\mu_f$ by a factor of two
up and down relative to their default values.
The second one is due to the fit errors in the LDMEs specified in
Table~\ref{tab:usedLDMEs}.
The third one is due to the lack of knowledge of the values of
$\langle\mathcal{P}^h(n)\rangle$ and the $\mathcal{O}(v^2)$ corrections to the
HQSS relations (\ref{eq:HQSSR}).
Both effects are estimated by evaluating Eq.~(\ref{eq:etacfact}) with
$\langle{\cal P}^h(n)\rangle=\xi m_c^2 \langle{\cal O}^h(n)\rangle$ and varying
$\xi$ in the range $-0.5<\xi<0.5$, so that $\xi$ is of order $v^2\approx0.23$
as obtained from potential model calculations \cite{Bodwin:2007fz}.

In Fig.~\ref{fig:etacplots}, the default NRQCD predictions are also broken down
to the individual Fock state contributions.
Evidently, the $h_c$ feeddown contribution is negligible owing to the small
${}^1\!P_1^{[1]}$ and ${}^1\!S_0^{[8]}$ SDCs, a feature that could not be
anticipated without explicit calculation, the more so as the $\chi_{cJ}$
feeddown contribution to prompt $J/\psi$ production is quite significant.
The most striking feature is, however, that the CSM, which is basically made up
just by the ${}^1\!S_0^{[1]}$ contribution, yields an almost perfect description
of the LHCb data, leaving practically no room for CO contributions.
While the ${}^1\!S_0^{[8]}$ and ${}^1\!P_1^{[8]}$ contributions comply with this
condition for all four $J/\psi$ LDME sets considered, the latter dictate a very
sizable ${}^3\!S_1^{[8]}$ contribution, which overshoots the LHCb data by up to
about one order of magnitude.
Even the LDME set that describes the LHCb data best, namely the one of
Ref.~\cite{Butenschoen:2011yh}, yields an unacceptable $\chi^2/\mathrm{d.o.f.}$
value of 257/7 with respect to the default NRQCD predictions. 
If we take the lower borders of the respective error bands in
Fig.~\ref{fig:etacplots} as a reference, then $\chi^2/\mathrm{d.o.f.}$ comes
down to 36.7/7, which is still very poor. 

In our second approach, we determine the $\eta_c$ and $h_c$ LDMEs without
recourse to the $J/\psi$ and $\chi_{cJ}$ LDMEs, by directly fitting the LHCb
data under certain simplifying assumptions.
First, we neglect the $h_c$ feeddown contributions by appealing to their
dramatic suppression in Fig.~\ref{fig:etacplots}.
Second, we neglect the ${}^1\!S_0^{[8]}$ and ${}^1\!P_{1}^{[8]}$ contributions
to direct $\eta_c$ production because of the $\mathcal{O}(v^4)$ suppression of
their LDMEs relative to the ${}^1\!S_0^{[1]}$ one, which is not compensated by
an inverse hierarchy in the respective SDCs.
In fact, the ${}^1\!S_0^{[8]}$ SDCs are only of the same order as the
${}^1\!S_0^{[1]}$ ones, while the ${}^1\!P_{1}^{[8]}$ ones are even smaller.
We are then left with the ${}^1\!S_{0}^{[1]}$ and ${}^3\!S_{1}^{[8]}$
contributions to direct $\eta_c$ production. 
As in Table~\ref{tab:usedLDMEs}, we include $\mathcal{O}(\alpha_s)$
corrections, but neglect $\mathcal{O}(v^2)$ corrections.
Our fitting procedure is as follows.
We first determine $\langle\mathcal{O}^{\eta_c}({}^1\!S_0^{[1]})\rangle$ from
the $\eta_c\to\gamma\gamma$ partial decay width \cite{Barbieri:1979be},
and then use it as input to fit
$\langle\mathcal{O}^{\eta_c}({}^3\!S_{1}^{[8]})\rangle$ to the LHCb data.
We are entitled to do so, since the difference between the CS LDMEs for
production and decay are of $\mathcal{O}(v^4)$ \cite{Bodwin:1994jh}.
In our determination of $\langle\mathcal{O}^{\eta_c}({}^1\!S_0^{[1]})\rangle$,
we set $\alpha=1/137$ and $\alpha_s(2m_c)=0.26$, and adopt the values
$\Gamma_{\eta_c}=(32.3\pm1.0)~\mathrm{MeV}$ and
$\mathrm{Br}(\eta_c\to\gamma\gamma)=(1.57\pm0.12)\times10^{-4}$
from Ref.~\cite{Agashe:2014kda}.
We thus obtain
$\langle\mathcal{O}^{\eta_c}({}^1\!S_{0}^{[1]})\rangle
=(0.24\pm0.02)~\mathrm{GeV}^3$, in reasonable agreement with the values of its
HQSS counterpart $\langle\mathcal{O}^{J/\psi}({}^3\!S_{1}^{[1]})\rangle$ in
Table~\ref{tab:usedLDMEs}, and
$\langle\mathcal{O}^{\eta_c}({}^3\!S_{1}^{[8]})\rangle
=(3.3\pm2.3)\times10^{-3}~\mathrm{GeV}^3$, yielding an excellent description of
the LHCb data, with $\chi^2/\mathrm{d.o.f.}=1.4/6$.
By HQSS, this provides an independent determination of
$\langle\mathcal{O}^{J/\psi}({}^1\!S_{0}^{[8]})\rangle=
\langle\mathcal{O}^{\eta_c}({}^3\!S_{1}^{[8]})\rangle$.
Observing that this value falls short of the lowest value in
Table~\ref{tab:usedLDMEs}, namely the one from Ref.~\cite{Butenschoen:2011yh},
by 6.47 standard deviations, we recover the striking disagreement encountered
in our first approach.
Such a low value of $\langle\mathcal{O}^{J/\psi}({}^1\!S_{0}^{[8]})\rangle$ is
in conflict with the ideas behind the high-$p_T$ fits in
Refs.~\cite{Chao:2012iv,Bodwin:2014gia}, which suggest a large
$\langle{\cal O}^{J/\psi}({}^1\!S_0^{[8]})\rangle$ value to render the
${}^1\!S_0^{[8]}$ contributions dominant in high-$p_T$ $J/\psi$
hadroproduction and to explain both the $J/\psi$ yield and polarization
observed experimentally.
However, unlike the $J/\psi$ case, the theoretical prediction of direct
$\eta_c$ hadroproduction is well under control.
In fact, there are no large NLO corrections in neither the CS or CO channels,
and the $h_c$ feeddown contributions are also small.

To summarize, we calculated, for the first time, the $\mathcal{O}(\alpha_s)$
corrections to the ${}^1\!S_0^{[1]}$ and ${}^1\!P_1^{[8]}$ SDCs as well as the
$\mathcal{O}(v^2)$ corrections to the ${}^1\!S_0^{[1]}$, ${}^1\!P_1^{[1]}$, and
${}^1\!P_1^{[8]}$ SDCs.
Using the $\eta_c$ LDMEs derived via HQSS from up-to-date $J/\psi$ LDMEs  
\cite{Butenschoen:2011yh,Chao:2012iv,Gong:2012ug,Bodwin:2014gia}, we
demonstrated that the CS contribution alone can nicely describe the new LHCb
data on prompt $\eta_c$ hadroproduction \cite{Aaij:2014bga}, while the full NLO
NRQCD predictions yield unacceptably large $\chi^2/\mathrm{d.o.f.}$ values, of
5.24 and above.
On the other hand, the CO contribution is almost exclusively exhausted by the
${}^3\!S_1^{[8]}$ channel, and the $h_c$ feeddown contribution is negligibly
small. 
This allowed us to directly fit
$\langle{\cal O}^{\eta_c}({}^3\!S_1^{[8]})\rangle$
to the LHCb data after determining
$\langle{\cal O}^{\eta_c}({}^1\!S_0^{[1]})\rangle$ from
$\Gamma(\eta_c\to\gamma\gamma)$, both in NRQCD through $\mathcal{O}(\alpha_s)$.
Conversion to $\langle{\cal O}^{J/\psi}({}^1\!S_0^{[8]})\rangle$ via HQSS
yielded a value that undershoots the expectation from the velocity scaling
rules by about one order of magnitude and the respective results from the NLO
NRQCD fits to $J/\psi$ production data currently on the market
\cite{Butenschoen:2011yh,Chao:2012iv,Gong:2012ug,Bodwin:2014gia} by at least
6.47 standard deviations.
Taking for granted that the LHCb results \cite{Aaij:2014bga} and the HQSS
relations (\ref{eq:HQSSR}) can be trusted and observing that the kinematic
region probed falls into mid-$p_T$ range, where neither large logarithms
$\ln(p_T^2/m_c^2)$ nor factorization breaking terms are expected, we are led to
conclude that either the universality of the LDMEs is in question or that
another important ingredient to current NLO NRQCD analyses has so far been
overlooked.

We are grateful to Sergey Barsuk and Maksym Teklishyn for providing us with
detailed information about the LHCb data \cite{Aaij:2014bga}.
This work was supported in part by
BMBF Grant No.\ 05~HT6GUA.

{\it Note added:} After submission, an alternative NRQCD analysis, at NLO in
$\alpha_s$, of prompt $\eta_c$ hadroproduction was reported \cite{Han:2014jya},
which finds the LHCb $\eta_c$ data \cite{Aaij:2014bga} to be consistent with a
2010 set of $J/\psi$ CO LDMEs \cite{Ma:2010yw} fitted to $J/\psi$ yield data
from CDF, in combination with an upper bound on
$\langle{\cal O}^{J/\psi}({}^1\!S_0^{[8]})\rangle$.
We wish to point out that this does not solve the notorious $J/\psi$
polarization puzzle.
In fact, this LDME set drives the polarization variable $\lambda_\theta$ in the
helicity frame to a positive value of approximately 0.4 at large $p_T$ and
central $y$ values (see first panel of Fig.~8 in \cite{Shao:2014yta} and second
panel of Fig.~2 in Ref.~\cite{Han:2014jya}), in disagreement with the Tevatron
and LHC measurements.

\end{document}